# CONSTRAINTS ON COSMOLOGICAL MODELS FROM ONCE AND FUTURE REDSHIFT SURVEYS[†]

MICHAEL S. VOGELEY

*Department of Physics and Astronomy, Johns Hopkins University*
*Baltimore, MD 21218 USA*
vogeley@pha.jhu.edu

## ABSTRACT

The power spectrum of density fluctuations measured from galaxy redshift surveys provides important constraints on models for the formation of large-scale structure. I present new results for a redshift sample of 15,000 galaxies, and review the limitations of current measurements. To span the decade of wavelength between the scales probed by galaxy surveys and COBE, measure the detailed shape of the power spectrum, and accurately examine the dependence of clustering on galaxy species, we require deeper samples with carefully controlled selection criteria and improved techniques for power spectrum estimation. I review plans for the Sloan Digital Sky Survey and describe a new method for estimating the power spectrum that optimally treats survey data with arbitrary geometry and sampling.

## 1. Introduction

The striking appearance of the galaxy distribution revealed by recent redshift surveys[1,2,3,4,5] suggests that the statistics of galaxy clustering provide a strong test of proposed models for the origin of structure in the universe. A variety of statistics have been applied to quantify the clustering of galaxies[6]. The power spectrum of density fluctuations and its Fourier conjugate, the autocorrelation function of the density field, describe the lowest order departures from homogeneity. In models with initially Gaussian density fluctuations, the power spectrum or autocorrelation function completely describes the inhomogeneities from which, in the standard class of models now considered, structure grows via gravitational instability.

The 3-D power spectrum has been measured for redshift samples of optical[7,8,9,10], infrared[11,12,13], and radio-selected[14] samples of galaxies. The power spectra all roughly follow a power law $P(k) \propto k^n$ with a slope ranging from $n \approx -2$ on small scales ($\lambda \lesssim 30h^{-1}$ Mpc) to $n \approx -1.1$ on intermediate scales ($30h^{-1}$ Mpc $< \lambda < 120h^{-1}$ Mpc). However, the power spectrum shape on larger scales and the overall normalization of the power spectrum differ among these samples. Some authors claim a turnover in the power spectrum[13]; others claim a continued rise[10] up to $200h^{-1}$Mpc. In addition, different groups find features in the power spectrum at different scales. Discrepancies among estimates may arise from differences in sample selection and/or the method of analysis, as well as from the uncertainty due to the finite volume of each survey.

---

[†]To appear in *Wide-Field Spectroscopy and the Distant Universe*, eds. S. J. Maddox and A. Aragón-Salamanca (World Scientific: Singapore); Proc. of 35th Herstmonceux Conference, Cambridge, U.K., July 4-8, 1994.

These redshift survey results, as well as angular correlations[15,16], indicate that the 'standard' $\Omega = 1$, $h = 0.5$ CDM model with biased galaxy formation lacks power on scales $\lambda > 50h^{-1}$ Mpc. More accurate measurements are now necessary to differentiate among the host of models that purport to match both the galaxy power spectrum and the microwave background anisotropy, e.g., mixed dark matter[17], 'tilted' inflationary models[18], and models with non-zero cosmological constant[19]. The measurements cited above do not have sufficient resolution to detect subtle features in the power spectrum and fail to provide constraints on the power spectrum on scales comparable to those probed by COBE. The dependence of clustering on galaxy species is also poorly understood.

Several ongoing and planned redshift surveys promise to improve this situation, and the development of new methods of analysis will allow better estimation from the extant and forthcoming data. Section 2 describes the problem of estimating the power spectrum and a method for doing so. Section 3 summarizes power spectrum results for a recently completed sample of 15,000 galaxies. Section 4 reviews results of recent surveys and discusses their limitations. Section 5 presents plans and predictions for the Sloan Digital Sky Survey. Section 6 describes new methods for power spectrum estimation. Section 7 presents conclusions.

## 2. Power Spectrum Estimation

The power spectrum measures the mean squared amplitude of each Fourier mode,

$$P(\mathbf{k}) = \langle |\delta(\mathbf{k})|^2 \rangle, \tag{1}$$

where

$$\delta(\mathbf{k}) = \frac{1}{V} \int \delta(\mathbf{x}) e^{i\mathbf{k}\cdot\mathbf{x}} d^3x \tag{2}$$

for a continuous density field $\rho(\mathbf{x})$ with density contrast $\delta(\mathbf{x}) = (\rho(\mathbf{x})/\langle\rho(\mathbf{x})\rangle) - 1$.

To estimate the power spectrum from a galaxy redshift survey, we must take into account the sampling density (determined by the magnitude limit) and geometry of the survey (determined by the angular coverage and depth). The sampling process introduces shot noise into the power spectrum. Because the observed power spectrum is a convolution of the true power with the Fourier transform of the spatial window function of the survey ($W(\mathbf{x}) = 1$ inside the survey and 0 outside),

$$P_{obs}(\mathbf{k}) = \int P_{true}(\mathbf{k}')|W(\mathbf{k}-\mathbf{k}')|^2 d^3k', \tag{3}$$

the survey geometry affects both the resolution of the measured power spectrum and the largest wavelength for which we obtain an accurate measurement.

An estimator for the power spectrum of a volume-limited galaxy sample is[10]

$$P(\mathbf{k}) = \left(\langle |\delta_{\mathbf{k}}|^2 \rangle - \frac{1}{N}\right) \left(\sum_{\mathbf{k}} |W_{\mathbf{k}}|^2\right)^{-1} \left(1 - |W_{\mathbf{k}}|^2\right)^{-1}, \tag{4}$$

where
$$\delta_{\mathbf{k}} = \frac{1}{N} \sum_j e^{i\mathbf{k}\cdot\mathbf{x_j}} - W_{\mathbf{k}}, \qquad (5)$$
is the Fourier transform of the galaxy positions minus the Fourier transform of the survey window. This method accounts for the contributions from shot noise (1/N term) and the survey geometry, and corrects the normalization (second factor) and shape (third factor) for the finite volume of the survey. Power spectrum estimates are usually shown (including all results below) for the angle-averaged quantity, $P(k)$.

The finite volume of a survey causes a power loss at large wavelengths because Eq. (5) implicitly assumes that the density of a sample equals the mean density of the universe. The third factor on the right side of Eq. (4) is an approximate correction for this large-scale power damping[14]. To correct more accurately for this effect, we compute the power spectrum for mock redshift surveys drawn from N-body simulations and compute the ratio of the mock survey power spectra to the true power spectrum. From this comparison we derive the correction to apply to the estimated power spectrum. The corrected power spectrum estimates are reliable up to scales where $|W_k|^2 \lesssim 0.2$.

We estimate the errors in the power spectrum caused by the finite survey volume and sampling density from the variance in the power spectra of 100 mock surveys drawn from an $\Omega h = 0.2$ CDM simulation. These Monte Carlo uncertainties include the effects of non-linear mode coupling and therefore are somewhat larger than those obtained by assuming purely Gaussian fluctuations[13].

## 3. The Power Spectrum of a 15,000 Galaxy Redshift Sample

We combine a recently completed survey of galaxies to $m_B \leq 15.5$ in the southern celestial hemisphere (SSRS2)[4] with the extension of the Center for Astrophysics redshift survey to $m_B \leq 15.5$ (CfA2)[2,20] to obtain a complete sample of $\sim 15,000$ galaxies over one-third of the sky. The large angular coverage and sampling density of this combined sample allows accurate estimation of the power spectrum on scales up to $\sim 200 h^{-1}$ Mpc[21].

This sample is large enough to allow us to test the reproducibility of our power spectrum results by direct comparison of the power spectrum obtained for the SSRS2, CfA2 North, and CfA2 South sub-samples, which probe different structures in comparable volumes. Figures 1a and 1b show the power spectra for each of these samples volume-limited to depths of $101 h^{-1}$Mpc and $130 h^{-1}$Mpc, respectively. The shapes of the power spectra are remarkably similar for $k > 0.2$. The reproducibility of the power spectrum for independent samples indicates that the power spectrum for optically-selected galaxies is robust for scales $\lambda \lesssim 100 - 150 h^{-1}$Mpc.

The power spectra of the 130-depth samples of both CfA2 North and SSRS2 exhibit an abrupt change of slope near $k \sim 0.2$. The reproducibility of this slope change in independent samples strengthens our suggestion that it is a real feature of the galaxy distribution, possibly reflecting the presence of voids $30 - 50 h^{-1}$ Mpc in diameter[2,4,5]. Detection of the $k \sim 0.2$ feature in a still larger volume is required to confirm a departure from CDM models.

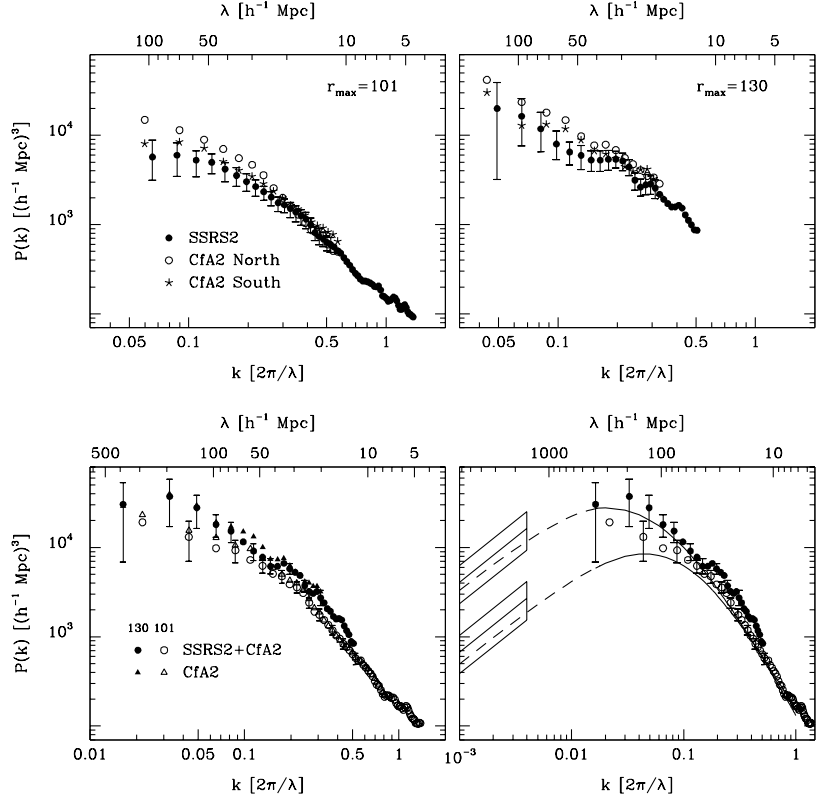

Fig. 1. POWER SPECTRA OF THE SSRS2+CfA2 SAMPLE. Upper left and upper right panels compare the power spectra of three volume-limited sub-samples, demonstrating the reproducbility of the power spectrum among independent volumes. Lower left panel shows the agreement between the combined SSRS2+CfA2 sample and the CfA2 sample alone. Lower right panel compares the power spectra of the combined SSRS2+CfA2 sample with limits on the mass power spectrum from COBE and the power spectra of two CDM models, all presented in *redshift space*. Upper solid line is the PS of CDM with $\Omega h = 0.2$ and $\Omega + \lambda_0 = 1$. Lower solid line is the PS of CDM with $\Omega h = 0.5$. Solid lines show the power spectra computed from N-body simulations of these models. Dashed lines show the PS of these models on linear scales. Both models are unbiased and normalized to $\sigma_8 = 1$. The boxes indicate constraints ($1\sigma$) on the redshift-space mass power spectrum from the COBE DMR for these two models.

In each sample, the overall amplitude of the power spectrum increases by a factor of ~1.4 from the $101h^{-1}$ Mpc to the $130h^{-1}$ Mpc samples. Luminosity segregation may cause amplitude variation in the power spectrum; the larger power spectrum amplitude corresponds to intrinsically brighter galaxies[10]. This possible effect should be taken into account in interpreting the results from flux-limited surveys.

The combined SSRS2+CfA2 sample yields our best constraints to date on the redshift-space power spectrum of optically-selected galaxies. Figure 1c shows results for CfA2 (combined CfA2 North and South[10] and the combined SSRS2+CfA2 sample. The power spectrum slope is $n \approx -2.1$ on scales up to $\lambda \sim 30h^{-1}$Mpc, then bends to $n \sim -1$ and continues to rise on scales up to $\lambda \sim 200h^{-1}$Mpc. At the largest scale where we compute the power spectrum for the deeper sample, $\lambda = 388h^{-1}$Mpc, the power spectrum appears to turn over. However, the uncertainty range for the power spectrum at this wavelength spans nearly an order of magnitude.

The redshift-space galaxy power spectrum and limits on the mass power spectrum at $z \sim 1000$ inferred from the COBE observations provide strong constraints on cosmological models. Here we bring the COBE data into the *redshift-space* of our power spectrum measurements and compare both CMB and galaxy observations with CDM models. In Figure 1d we plot our best estimate of the power spectrum for optical galaxies (SSRS2+CfA2) along with error boxes for the mass power spectrum that obtain from the COBE DMR experiment for the two CDM models presented[19,22] with $Q_{rms} = 17.1 \pm 2.9 \mu K^{23}$. The amplification of the power spectrum in redshift space is described by Kaiser[24]. We plot redshift-space mass power spectrum (identical to the galaxy power spectrum if unbiased) for CDM with $\Omega = 1$, $h = 0.5$ and CDM with $\Omega = 0.4$, $h = 0.5$, and $\Omega + \lambda_0 = 1$, both normalized to $\sigma_8 = 1$. We compute these power spectrum from particle-mesh N-body simulations. With this normalization, these CDM models require no biasing of galaxies vs. mass. Unbiased CDM with $\Omega = 1$ lacks power on scales $\lambda \sim 100h^{-1}$ Mpc. CDM with $\Omega h = 0.2$, $\Omega + \lambda_0 = 1$, and $b \approx 1$ ($\pm 0.2$ for the $1\sigma$ COBE uncertainty) is consistent with both the galaxy power spectrum and COBE. A somewhat better fit would obtain for slightly larger $\Omega h^{19}$. On scales sampled by the galaxy power spectrum, the power spectrum of an $\Omega h \sim 0.2$ model with $\lambda_0 = 0$ is nearly identical to that of a model with $\lambda_0 = 1 - \Omega$: either type of model is consistent with the SSRS2+CfA2 power spectrum. However, a non-zero cosmological constant strongly affects the the amplitude of mass fluctuations at the present epoch implied by COBE[19].

## 4. Limitations of Current Measurements

Figure 2 shows the power spectrum of the SSRS2+CfA2 sample[21], the IRAS 1.2Jy survey[12], the QDOT survey of IRAS galaxies[13], and the power spectrum inverted from the angular correlation function of the APM catalog[16]. The shapes of the two SSRS2+CfA2 samples and the IRAS 1.2Jy sample are consistent within the errors. However the amplitudes of these spectra differ quite significantly, suggesting that the clustering amplitude of different species extends over the full range of wavelength scales. The QDOT survey exhibits a large excess in power on scales $\lambda > 75h^{-1}$ Mpc over the shallower but more densely sampled 1.2Jy IRAS sample. One interpretation of this result is that the clustering amplitude increases with depth within the volume of the QDOT sample[13].

The APM curve in Figure 2 is an estimate of the *real-space* power spectrum. The corresponding redshift-space spectrum would be steeper than this, and therefore in disagreement with the other optically-selected power spectra plotted here (SSRS2+CfA2). If the clustering amplitude is an increasing function of galaxy luminosity, as suggested by the results in section 3, then the power spectrum inferred from angular correlations (which, by definition, use an apparent-magnitude limited sample) will have too steep a slope because the clustering amplitude increases with depth and thus with wavelength scale.

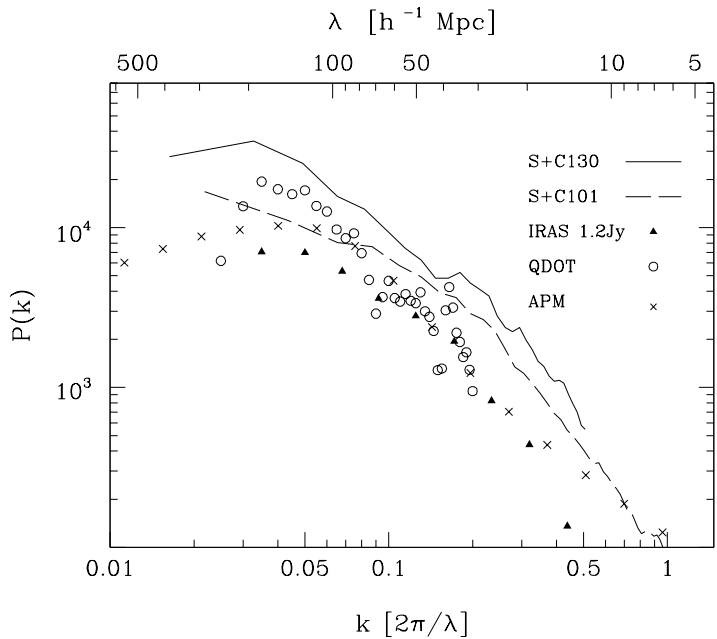

Fig. 2. POWER SPECTRA OF DIFFERENT GALAXY SURVEYS. Solid and dashed curves show redshift-space power spectra of volume-limited samples of the optically-selected SSRS2+CfA2 sample. Triangles and circles show the redshift-space power spectra of flux-limited samples of the 1.2Jy and QDOT (1/6 of the 0.6Jy sample) IRAS surveys. Crosses show the real-space power spectrum inferred from the angular correlations of the APM catalog.

Despite the detailed differences, the 'standard' CDM model is strongly challenged, and models with a larger ratio of large to small-scale power favored, by all of these results. More stringent tests of models require that we (1) quantify the dependence of galaxy clustering on galaxy species, (2) extend our knowledge of the galaxy power spectrum to overlap with the scales probed by COBE, and (3) estimate the power spectrum with sufficient resolution to detect or rule out the presence of features predicted by some models.

The present situation is unsatisfying because a synthetic interpretation of these data requires assumptions about the relative clustering and velocity fields of different galaxy species[25]. Ideally, we would resolve this uncertainty by obtaining a sample of galaxies for which we can directly examine these dependences, i.e., we select the different sub-samples from within the same physical volume. Such analyses require more detail (galaxy colors, morphology, spectral type, etc.) and a larger number of galaxies (so that sub-samples are not too small to analyze) than is generally available for existing surveys.

The CfA, SSRS, and IRAS redshift surveys dramatically reveal the nature of galaxy clustering in the nearby universe, but these surveys are still relatively small compared with the scale probed by COBE. The finite volume of these surveys limits both the largest wavelength scale and the resolution with which we can accurately estimate the power spectrum. Estimation of the uncertainty of the power spectrum due to the finite volume probed is difficult because this uncertainty depends on the assumptions about the spectrum itself as well as the behavior of higher order moments[13].

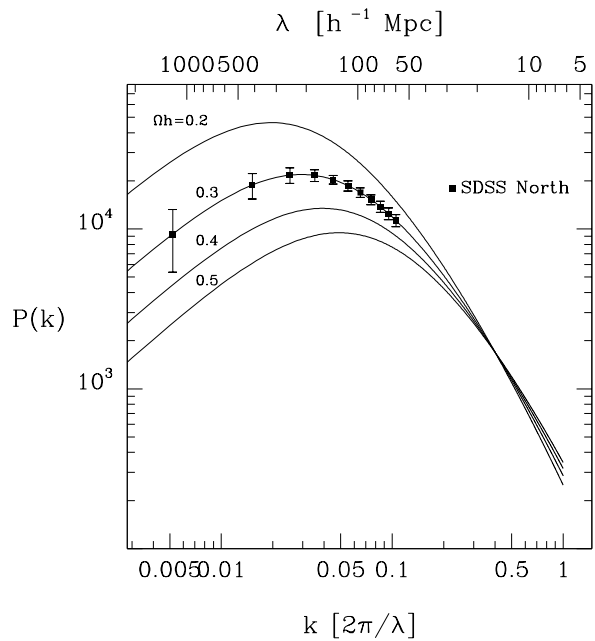

Fig. 3. UNCERTAINTY OF THE SDSS POWER SPECTRUM. The $1\sigma$ uncertainty expected for a volume-limited (to $M^*$) sample of the SDSS northern redshift survey, assuming Gaussian fluctuations and a $\Omega h = 0.3$ CDM model, compared to power spectra for CDM with different $\Omega h$. Error bars on smaller scales are of similar or smaller size than the symbols.

The turnover in the power spectrum seen in Figure 2 at $k \sim 0.03$ is uncertain and the resolution of the power spectrum near the peak of the galaxy power spectrum, which corresponds to the scale entering the horizon at the time of matter-radiation equality, is too poor to detect subtle features such as those produced in, e.g., models with a large baryonic mass component[26].

Deep pencil beam surveys[27] probe $\sim 1000 h^{-1}$ Mpc scales, but the narrow geometry allows leakage of power from smaller wavelength scales, complicating the interpretation of the 1-D power spectrum. Another approach to probing larger scales is to sparsely sample the galaxy distribution, e.g., the 1/6 sampling of IRAS galaxies in the QDOT survey. If the true fluctuations are Gaussian, then such an approach is quite useful[28]. However, sparse sampling comprises the utility of such surveys for examining higher order statistics (and thus for testing the Gaussian hypothesis).

## 5. Plans and Predictions for the SDSS

Several ongoing surveys, including the Las Campañas[5], IRAS PSC[29], ESO key project[30], and Century[31] surveys, are close to yielding new measurements of the power spectrum. Still larger planned surveys, the Sloan Digital Sky Survey (SDSS hereafter) and AAT 2dF survey, will yield measurements an order of magnitude more accurate than currently possible. Here I review plans for the SDSS and describe the constraints on the galaxy power spectrum expected for this survey.

The SDSS will obtain precision photometry in five bandpasses $(u', g', r', i', z')$ over one-quarter of the sky, centered at the North Galactic Cap. In parallel with the imaging survey, spectroscopy will be obtained for $10^6$ galaxies and a smaller number of quasar candidates and stars. These galaxies will be selected *after* correcting for Galactic extinction, in order to yield the largest possible volume-limited samples. To the approximate magnitude limit of $g' = 18.3$, the median redshift of the spectroscopic sample will be $\sim 300 h^{-1}$ Mpc. Thus the SDSS redshift survey will encompass a volume that is two orders of magnitude larger than any current sample. In addition, deeper imaging and spectroscopy of a carefully chosen fraction of the southern galactic hemisphere will be obtained.

In models for the formation of large-scale structure that are consistent with COBE, the amplitude of galaxy density fluctuations on large scales could be smaller than the uncertainties in the selection function for current surveys. Therefore, a requirement for the accuracy of the photometry, the reliability of the spectroscopic target selection, and the precision of the spectroscopy is to produce a galaxy sample for which the accuracy of clustering measures is limited by statistical (e.g., due to the volume sampled) rather than systematic uncertainty. Clearly the most demanding of these is the requirement for photometric consistency and accuracy over one-quarter of the sky; current surveys are limited by the poor sensitivity, non-linearity, and difficulty of calibration of photographic plates over more than a small area on the sky. Thus, the digital imaging survey is an essential prerequisite for the redshift survey.

A further benefit of the detailed imaging and high-resolution spectroscopy is that these data will allow classification of galaxies using several methods, e.g., by color, morphological type, and spectral type. Thus, this sample will help to resolve the uncertainties in the dependence of clustering on galaxy type described above.

The SDSS redshift survey volume (1) is large enough to include many independent structures on the scale of the "Great Wall," and (2) has sufficient angular coverage and depth to ensure that measurement of the fluctuation spectrum on scales of a few hundred $h^{-1}$ Mpc is not strongly affected by aliasing. The very sharp Fourier window function will allow precise measurement of the power spectrum over a large range of wavelength scales, and allow tests for features like those described in section 5.

The precision with which the northern spectroscopic galaxy sample of the SDSS will allow measurement of the power spectrum, assuming the above criteria for the survey, is illustrated in Figure 3, which shows the power spectrum of the $\Omega h = 0.3$ cold dark matter model (which is roughly consistent with existing data), together with an estimate of the uncertainty that we expect for the SDSS northern redshift sample, computed assuming Gaussian fluctuations[13]. (Compare these uncertainties with those plotted in Figure 1.) If this extrapolation of current measurements to larger scales is roughly correct, then the SDSS will accurately probe the galaxy power spectrum beyond the turnover scale, and up to the scales probed by COBE. The northern and southern survey regions will provide independent measurements of the power spectrum, and the combined sample will yield constraints on a wavelength scale twice the depth of the spectroscopic survey.

The first year of operation will yield redshifts for $2 \times 10^5$ galaxies, an order of magnitude larger than any existing survey. An appropriate choice of survey strategy will allow us to obtain excellent constraints on the power spectrum on large wavelength scales with these data alone. An example of a reasonable compromise between the demands of imaging in driftscan mode, the desire to efficiently tile the survey region with spectroscopic fields, and a statistically optimal geometry for one year of observing with this telescope would be a set of five slices spread over the north galactic cap, each slice having an area of approximately $5° \times 120°$. Because the intervening unobserved slices provide nearly redundant information regarding clustering on scales larger than their separation, the first year's data alone will be sufficient to constrain the power spectrum on the COBE scale with uncertainty only slightly larger than for the whole survey. We will apply novel techniques, such as those described below, to analyze these data in such a way as to minimize the aliasing that might otherwise enter due to the narrow declination range of the individual slices.

## 6. New Methods for Power Spectrum Estimation

The standard methods for power spectrum estimation (cf. section 2) work reasonably well for data in a large, contiguous, three-dimensional volume, with homogeneous sampling of the galaxy distribution. However, these methods are less efficient when applied to data in oddly-shaped and/or disjoint volumes, or when the sampling density of galaxies varies greatly over these regions. In addition, because the measured power spectrum is a convolution of the true power with the survey window function, the power in different modes can be highly coupled. In other words, plane waves do not form an optimal eigenbasis for expansion of the the galaxy density field sampled by redshift surveys. We desire methods for power spectrum estimation that optimally weight the data in each region of the survey, taking into account our prior knowledge of the nature of the noise and clustering in the galaxy distribution.

### 6.1. The Cross-Correlation Power Spectrum

One problem that arises when one combines disjoint survey regions is that the window function (in the Fourier domain) of each region typically has very wide sidelobes, thus the power spectrum of the full data set is plagued by aliasing. However, nearly all of the sensitivity to large wavelength fluctuations arises from comparing the density in different regions, not from the fluctuations within each region[32]. The window function for the whole survey is the sum of the individual window functions,

$$W_{tot}(\mathbf{k}) = \sum_i W_i(\mathbf{k}), \qquad (6)$$

and the Fourier transform of the full sample is the sum of the transforms of the subregions (cf. Eq. 5),

$$\delta_{tot}(\mathbf{k}) = \sum_i [\delta_i(\mathbf{k}) - W_i(\mathbf{k})] \qquad (7)$$

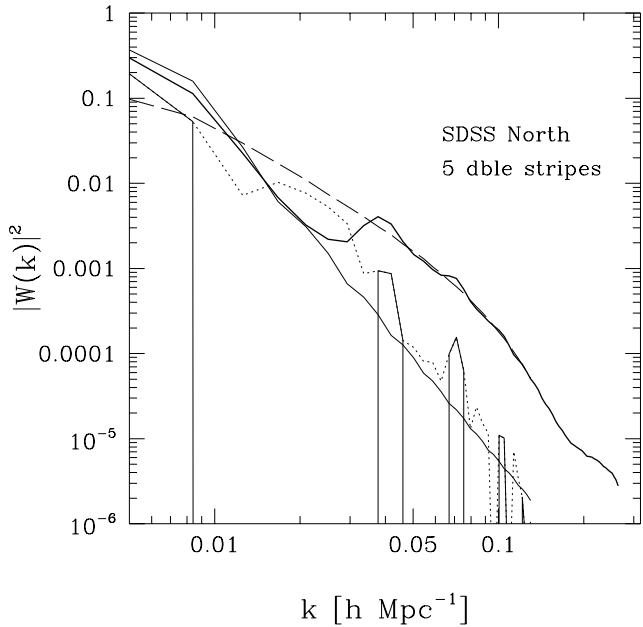

Fig. 4. COMPONENTS OF THE FOURIER WINDOW FUNCTION FOR THE SDSS. Lower (approx. straight) solid line depicts the angle-average window function for a volume-limited (to $M^*$) sample of the SDSS northern redshift survey. Upper (heavy) solid line shows the window function for a sub-sample of five $5° \times 120°$ slices, spread within the SDSS region, which is the sum of the autocorrelation window function for the individual slices (long-dashed line), and the window function of the cross-correlations between the disjoint slices (lower light solid line; dotted line shows the absolute value of the negative regions of this function, which is not positive definite). Use of only the cross-correlation component of the 'slice' power spectrum avoids the strong aliasing caused by the sidelobes of the autocorrelation part.

The power spectrum for the full survey is

$$\begin{aligned} P(\mathbf{k}) &= \delta(\mathbf{k})\delta^*(\mathbf{k}) \\ &= \sum_i \delta_i(\mathbf{k})\delta_i^*(\mathbf{k}) + \sum_{i>j} \left[\delta_i(\mathbf{k})\delta_j^*(\mathbf{k}) + \delta_i^*(\mathbf{k})\delta_j(\mathbf{k})\right]. \end{aligned} \qquad (8)$$

The first term in Eq. (8) measures the autocorrelation of the individual regions and the second term measures only the cross-correlation of different regions. The window function for this power spectrum can likewise be broken down into autocorrelation and cross-correlation terms,

$$|W(\mathbf{k})|^2 = \frac{1}{\sum_i V_i^2} \left[\sum_i V_i^2 |W_i(\mathbf{k})|^2 + \sum_{i>j} V_i V_j \left\{W_i(\mathbf{k})W_j^*(\mathbf{k}) + W_i^*(\mathbf{k})W_j(\mathbf{k})\right\}\right]. \qquad (9)$$

Figure 4 shows the total, auto, and cross-correlation window functions for the SDSS 'first year' strategy described above. We see that the autocorrelations cause the wide sidelobes in the window function of the full survey. Thus we can obtain a much cleaner estimate of the power on large wavelength scales by using only the "cross-correlation power spectrum,"

$$P_{cross}(\mathbf{k}) = \sum_{i>j}[\delta_i(\mathbf{k})\delta_j^*(\mathbf{k}) + \delta_i^*(\mathbf{k})\delta_j(\mathbf{k})]. \tag{10}$$

Note that the peaks in the cross-correlation window occur because the slices in this example are regularly spaced by 20°. We can reduce these peaks, and thus obtain a more compact window, if we irregularly space the slices.

*6.2. Eigenmode Analysis of Redshift Surveys*

The previous discussion demonstrates the gains that obtain from more carefully combining different regions of the survey volume. The cross-correlation power spectrum method effectively downweights the fluctuations within each region in favor of the fluctuations between disjoint regions. Here we generalize this procedure in order to find an optimal set of spatial filters to probe the density fluctuations.

Rather than directly compute the Fourier transform of the distribution of objects, we expand the observed density field in the natural orthonormal basis which obtains for each survey using our prior knowledge of the survey geometry, selection function, and clustering of galaxies, and find the most likely power spectrum model in a Bayesian fashion. Expansion of the observed density field in this basis is known as the Karhunen-Loève transform[33].

Dividing the survey volume into cells $V_i$, we compute the correlation matrix of expected counts

$$\begin{aligned} R_{ij} &= \langle N_i N_j \rangle \\ &= \langle N_i \rangle \langle N_j \rangle (1 + \langle \xi \rangle_{ij}) + \delta_{ij} N_i, \end{aligned} \tag{11}$$

where $\delta_{ij} = 0$ for $i \neq j$, $N_i$ is the galaxy count in the $i$th cell, and

$$\langle \xi \rangle_{ij} = \frac{1}{V_i V_j} \int \xi(\mathbf{x_i} - \mathbf{x_j}) dV_i dV_j. \tag{12}$$

We assume a model for $\xi$ that is consistent with previous observations. The column vectors $\mathbf{\Psi}_j$ of the unitary tranformation that diagonalizes the correlation matrix are the eigenfunctions of the density field of the survey (thus solving the eigenvalue equation $\mathbf{R} \cdot \mathbf{\Psi}_j = \lambda_j \mathbf{\Psi}_j$).

Figure 5 shows the first 12 density eigenmodes for the geometry, selection function, and correlation function of the first CfA slice[1]. Figure 6 shows the Fourier transform of these eigenmodes. As we increase the volume of the survey, the Fourier windows of the eigenmodes sharpen; in the limit of an infinite or periodic volume, the eigenmodes are plane waves and we recover the Fourier expansion.

Thus, we expand the observed counts in this orthonormal basis: $N_i = \beta^j \Psi_{ij}$ (Einstein summation convention), which defines the transform $\beta^j = \Psi^{ij} N_i$. Sorting these functions by decreasing eigenvalue $\lambda$ yields the set of eigenfunctions in order of decreasing signal to noise. Figure 7 shows the expected power per mode of the Karhunen-Loève transform analogous to the power spectrum of the Fourier expansion. The total power per mode is the sum of the true clustering power, shot noise, and the mean density (for an infinite or periodic volume the $n = 1$ K-L mode corresponds to the $k = 0$ Fourier mode).

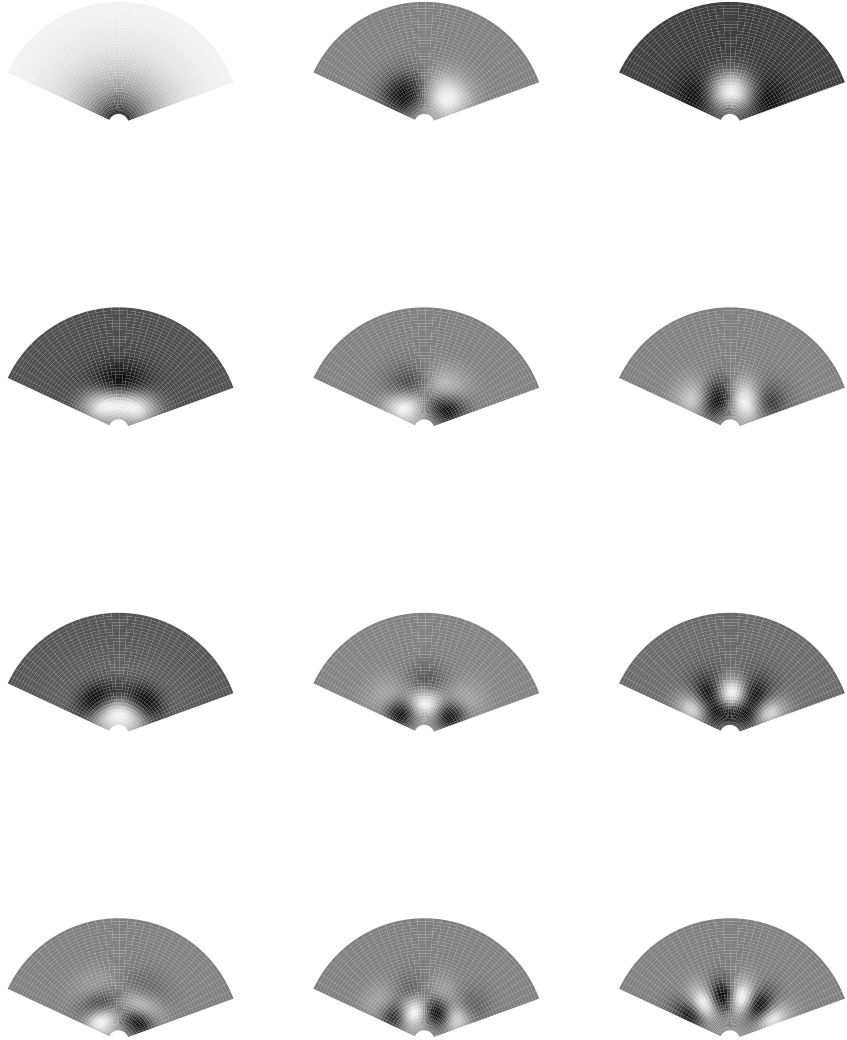

Fig. 5. EIGENMODES OF THE CfA SLICE. The Karhunen-Loève transform uses the geometry, selection function, and known clustering of galaxies to find the unique statistically orthogonal basis set for the density fluctuations in a particular survey. Here we plot the twelve most significant eigenfunctions for the first CfA slice.

Because the $\beta^j$ are statistically orthogonal and because we can easily compute the expectation value and variance of the power per eigenmode for any power spectrum

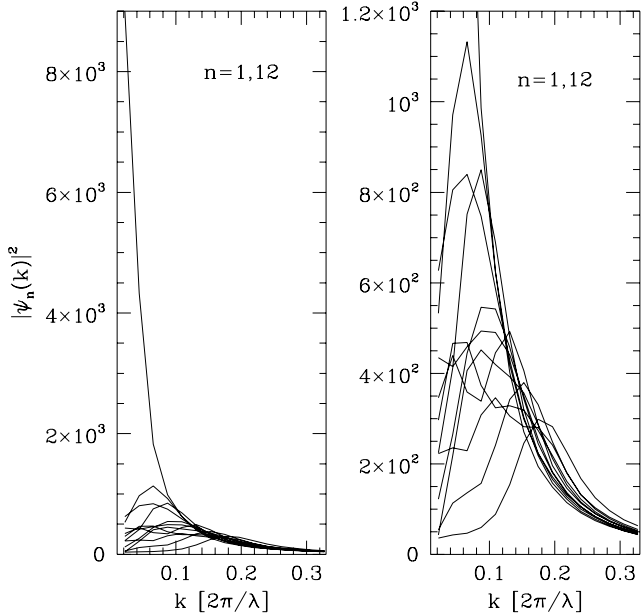

Fig. 6. FOURIER WINDOW FUNCTIONS OF THE CFA SLICE EIGENMODES. Similar to the window functions plotted in Figure 4, the modulus-squared of the Fourier transform of the K-L eigenmodes shows that each eigenmode samples a limited range of wavenumber. The spike at $k = 0$ in the left hand panel is the window function of the K-L mode that carries most of the information about the mean density. The right panel expands the left to show the shape of the window functions of the eigenmodes plotted in Figure 5.

model
$$\langle \beta_j^2 \rangle = \mathbf{\Psi}_j^{-1} \cdot \mathbf{R}^{model} \cdot \mathbf{\Psi}_j, \tag{13}$$
hypothesis testing is a straightforward process. Note that this method requires an initial guess at the power spectrum, but the form of the eigenfunctions does not depend sensitively on this assumption, and we can easily iterate the process. Using this technique, we hope to extract new constraints on the power spectrum on large wavelength scales from data that heretofore proved difficult to analyze because of their odd geometry, e.g., deep pencil beam surveys.

## 7. Conclusions

Statistical measures of the large-scale structure revealed by redshift surveys of the nearby universe like the CfA, SSRS, and IRAS surveys, together with measurement of the CMB anisotropy by COBE, successfully narrow the range of acceptable theoretical models, ruling out, for example, the previously 'standard' CDM model, and suggest consideration of several new models. The discriminatory power of galaxy clustering statistics and the increasing predictive power of theoretical models prompt deeper redshift surveys (e.g., Las Campañas, AAT 2dF, SDSS, DEEP[34]) and the development of more sophisticated methods of analysis (e.g., the K-L transform) in order to fill the gaps in our knowledge of large-scale structure that surveys of the nearby universe are unable to answer. The next few years of observational cosmology will be exciting: with larger surveys and improved techniques, we will explore the decade of wavelength scales between current observations and the scales probed by COBE, probe the fluctuation spectrum with higher resolution, examine the detailed dependence of clustering on galaxy species, and begin to explore the evolution of large-scale structure.

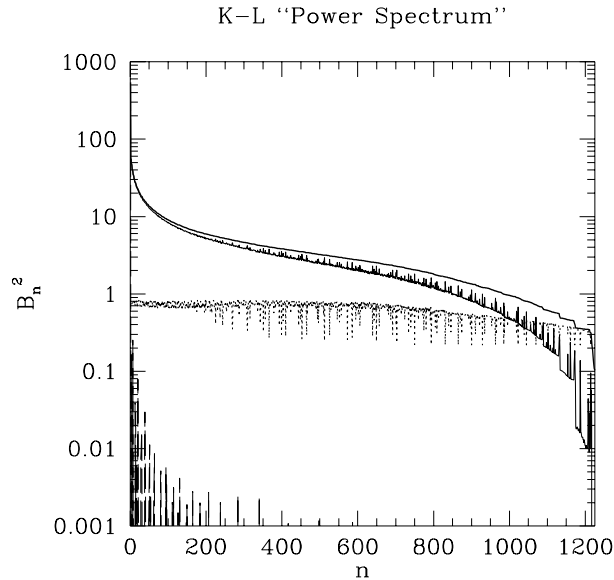

Fig. 7. POWER PER MODE OF THE K-L EXPANSION. The upper solid line plots the eigenvalue for each mode, which is the expectation value of the total power per mode. Successively lower lines show the contributions to the power from galaxy clustering, shot noise, and the mean density.

## Acknowledgements

I thank the many individuals who contributed to the work summarized above: Luiz da Costa, Margaret Geller, John Huchra, and Changbom Park for the analyses of the SSRS2 and CfA2 surveys; Alex Szalay for the development of improved methods for power spectrum estimation; Alex Szalay, Andy Connolly, David Weinberg, and Michael Strauss for the discussion of measurements of large-scale structure with the SDSS. I acknowledge partial support from the Sloan Digital Sky Survey and NSF grant 9020380.